
\magnification \magstep1
\raggedbottom
\openup 2\jot
\voffset6truemm
\headline={\ifnum\pageno=1\hfill\else
\hfill{\it Lagrangian theory of constrained systems:
cosmological application}\hfill \fi}
\centerline {\bf LAGRANGIAN THEORY OF CONSTRAINED SYSTEMS:}
\centerline {\bf COSMOLOGICAL APPLICATION}
\vskip 1cm
\centerline {\bf Giampiero Esposito$^{1,2}$,
Gabriele Gionti$^{3}$, Giuseppe Marmo$^{1,2,4}$}
\centerline {\bf and Cosimo Stornaiolo$^{1,2}$}
\vskip 1cm
\centerline {{\it ${ }^{1}$Istituto Nazionale di Fisica Nucleare}}
\centerline {\it Mostra d'Oltremare Padiglione 20, 80125 Napoli, Italy;}
\centerline {{\it ${ }^{2}$Dipartimento di Scienze Fisiche}}
\centerline {\it Mostra d'Oltremare Padiglione 19, 80125 Napoli, Italy;}
\centerline {{\it ${ }^{3}$Scuola Internazionale Superiore di Studi
Avanzati}}
\centerline {\it Via Beirut 2-4, 34013 Trieste, Italy;}
\centerline {{\it ${ }^{4}$Departamento de Fisica Teorica}}
\centerline {\it Facultad de Ciencias Fisicas}
\centerline {\it Universidad Complutense, 28040 Madrid, Spain.}
\vskip 1cm
\noindent
{\bf Summary. -}
Previous work in the literature has studied the Hamiltonian structure of
an $R^{2}$ model of gravity with torsion in a closed
Friedmann-Robertson-Walker universe. Within the framework of Dirac's
theory, torsion is found to lead to a second-class primary constraint
linear in the momenta and a second-class secondary constraint quadratic in
the momenta.

This paper studies in detail the same problem at a
Lagrangian level, i.e. working on the tangent bundle
rather than on phase space. The corresponding analysis
is motivated by a more general program, aiming to
obtain a manifestly covariant, multisymplectic
framework for the analysis of
relativistic theories of gravitation regarded
as constrained systems. After an application of the
Gotay-Nester Lagrangian analysis, the paper deals
with the generalized method, which has the advantage
of being applicable to
any system of differential equations in implicit form.
Multiplication of the second-order Lagrange equations by a
vector with zero eigenvalue for the Hessian matrix
yields the so-called first-generation constraints.

Remarkably, in the cosmological model here considered,
if Lagrange equations are studied using second-order
formalism a second-generation constraint is found which
is absent in first-order formalism. This happens since
first- and second-order formalisms are inequivalent.
There are, however, no {\it a priori} reasons for arguing that
one of the two is incorrect. First- and second-generation
constraints are used to derive physical
predictions for the cosmological model.
\eject
\leftline {\bf 1. - Introduction.}
\vskip 0.3cm
In recent years part of the theoretical-physics
community has started appreciating that constrained systems
may be also studied using a tangent-bundle
formalism [1], rather than working on phase space.
{}From the point of view of relativistic theories of gravitation,
this approach appears relevant since it may lead to a fully
covariant treatment of the gravitational field regarded as
a constrained system, without having to use a 3+1 split
of the Lorentzian space-time manifold.

Relying on previous work on
cosmology in [2], we perform the analysis, in  a
closed Friedmann-Robertson-Walker (hereafter referred to as
FRW) universe, of a model of gravity with non-vanishing
torsion where the Lagrangian density is proportional to the
square of the scalar curvature. Of course, this is only a toy
model, but its analysis deserves careful consideration because
if one tries to quantize general relativity (GR) or theories with
non-vanishing torsion within the framework of perturbative
renormalization, one finds that quantized GR is perturbatively
non-renormalizable, and in both theories the effective action
acquires terms quadratic in the full Riemann-curvature tensor,
i.e. terms proportional to $R_{abcd}R^{abcd}$, $R_{ab}R^{ab}$,
${\Bigr(R_{\; \; a}^{a}\Bigr)}^{2}$. In the most general case,
these theories, which are non-linear in the curvature, are either
non-unitary or perturbatively non-renormalizable [3].
Moreover, they are much studied as gauge
theories of gravitation [4-6], although the corresponding Cauchy
problem at the classical level may be ill-posed.
However, to improve our understanding of all difficulties and
limits of the perturbative-renormalization program in quantum
gravity, and to check the consistency or inconsistency of
inflationary-universe models based on these non-linear
theories, it appears essential to complete a detailed analysis
of these models of gravity.

In the torsion-free case, a relevant cosmological application
was studied, for example, in [7], where an action
functional was considered which is purely quadratic in the
trace of the Ricci tensor. The motivation for that
quantum-cosmological analysis was to obtain a Wick-rotated
path integral whose integrand does not blow up exponentially
after suitable conformal rescalings of the 4-metric. As a first
step of the program motivated so far, a similar model was
proposed in [2] (see corrections in [8]),
from a Hamiltonian point of view,
by applying Dirac's theory of constrained Hamiltonian systems
at the classical level. The resulting idealized cosmological
model is a constrained system with second-class constraints,
which arise, as shown below, by virtue of torsion.
In the following part of this introductory
paragraph, to help the reader, we summarize the
Hamiltonian analysis appearing in [2].

In our FRW model, spatial homogeneity and isotropy
imply that, in a coordinate frame, the only non-vanishing components
of the torsion tensor are
$
S_{10}^{\; \; \; 1}=S_{20}^{\; \; \; 2}=S_{30}^{\; \; \; 3}=Q(t).
$
Denoting by $N$ the lapse function, by $a(t)$
the 3-sphere radius of the closed FRW metric, and defining $\alpha \equiv
\log(a)$, ${Q\over N} \equiv y$, $\tau \equiv \int N \; dt$, $' \equiv
{d\over d\tau}$, the action functional
$$
I \equiv \kappa \int {\Bigr(R_{\; \; a}^{a}\Bigr)}^{2}
\; \sqrt{-{\rm det} \; g} \; d^{4}x
\; \; \; \; ,
\eqno (1.1)
$$
after integration on the 3-sphere,
takes the form $I= \int {\widetilde L} \; d\tau$, where
$$
{\widetilde L} \equiv \mu e^{3\alpha}
{\biggr[e^{-2\alpha}+2(\alpha')^{2}-12y\alpha'+16y^{2}
+\alpha''-2y' \biggr]}^{2}
\; \; \; \; ,
\eqno (1.2)
$$
and $\mu$ is a proportionality constant, denoted by $\mu^{0}$ in
equation (2.13) of [2].
Note that the point Lagrangian (1.2) is obtained by inserting
the FRW hypothesis into the field-theory Lagrangian of (1.1).
By a direct calculation, one can check that this procedure
is correct in the particular
case of FRW cosmologies in general relativity [9].
For alternative theories, such a method was found to be
correct in [7] in the FRW torsion-free case. We are thus
using the assumptions about the FRW symmetry to reduce
ourselves to the study of point Lagrangians without having
to build the complete Hamiltonian treatment for a generic
curvature-squared field theory with torsion.

Since the addition to the
Lagrangian of a total derivative leads to an equivalent set of
field equations, we use this property to cast the theory in
Hamiltonian form. To eliminate the square of second-order
derivatives appearing in $\widetilde L$ one is thus led to
define [2,7,10]
$$
L \equiv {\widetilde L} - {d\over d\tau}\Bigr[(\alpha'-2y)z\Bigr]
\; \; \; \; ,
\eqno (1.3)
$$
where $(\alpha'-2y)$ is proportional to the trace of the
extrinsic-curvature tensor of the 3-sphere, and $z$ is
obtained differentiating $\widetilde L$ with respect to the
highest derivative, i.e. defining $z \equiv
{\partial {\widetilde L}\over \partial x''}$, where
$x \equiv \alpha -2 \int y \; d\tau$. Thus, setting
$y \equiv u'$, the Lagrangian defined in equation (1.3)
becomes (cf. equation (2.1))
$$
L = 16z(u')^{2}+2z(\alpha')^{2}-12zu'\alpha'+2u'z'
-z'\alpha'+ze^{-2\alpha}-{z^{2}\over 4\mu}e^{-3\alpha}
\; \; \; \; .
\eqno (1.4)
$$
For comments on this choice of variables, see again [2,7].
Hence, defining $p_{\alpha} \equiv {\partial L \over
\partial \alpha'}$,
$p_{u} \equiv {\partial L \over \partial u'}$,
$p_{z} \equiv {\partial L \over \partial z'}$
(cf. section 2), one finds the primary constraint
$\phi_{1} \approx 0$, where the weak-equality symbol
$\approx$ denotes an equality which only holds on the constraint
manifold [11-13], and
$$
\phi_{1} \equiv 2p_{\alpha}+p_{u}-4zp_{z}
\; \; \; \; .
\eqno (1.5)
$$
The corresponding effective Hamiltonian $\widetilde H$ on the
whole phase space is given by
${\widetilde H} \equiv H_{c}+\gamma
\phi_{1}$, where $H_{c}$, the Legendre transform of $L$, takes
the form
$$
H_{c}=-4zp_{z}^{2}+{p_{u}p_{z}\over 2}
+{z^{2}e^{-3\alpha}\over 4\mu}-ze^{-2\alpha}
\; \; \; \; .
\eqno (1.6)
$$
The constraint $\phi_{1}$ is preserved in time by requiring that
its Poisson bracket with $\widetilde H$, as defined in [11],
should vanish: $\Bigr \{\phi_{1},{\widetilde H} \Bigr \}
\approx 0$. This leads to the secondary constraint
$$
\phi_{2} \equiv 16zp_{z}^{2}-2p_{u}p_{z}+{7z^{2}\over 2\mu}
e^{-3\alpha}-8ze^{-2\alpha}
\; \; \; \; .
\eqno (1.7)
$$
The constraints $\phi_{1}$ and $\phi_{2}$ are second-class, since,
after evaluating their Poisson bracket, one obtains a function
which does not vanish when $\phi_{1}$ and $\phi_{2}$ are set to zero.
Hence $\gamma$ can be obtained as
$\gamma = -{{\Bigr \{\phi_{2},H_{c} \Bigr \}}\over
{\Bigr \{\phi_{2},\phi_{1} \Bigr \}}}$ as in equations
(2.24)-(2.25) of [2]. Note that, since $\phi_{1}$
and $\phi_{2}$ are second-class, we can define a new Poisson
bracket, the so-called Dirac brackets, in which second-class
constraints can be set strongly to zero [11-13],
i.e. they behave as Casimir functions for Dirac
brackets. This implies that the canonical Hamiltonian $H_{c}$ in equation
(1.6) may be also written as
$$
H_{c}=-2zp_{z}^{2}-p_{\alpha}p_{z}+{z^{2}e^{-3\alpha}\over 4\mu}
-ze^{-2\alpha}
\; \; \; \; ,
\eqno (1.8)
$$
which formally coincides with the torsion-free result [7].
The effective Hamiltonian $\widetilde H$,
however, is not a linear combination of constraints and hence
does not vanish in general.
One then finds the field equations (2.27)-(2.32) of [2].

After this introductory paragraph, we can summarize the plan
of our paper as follows. Section 2 performs a Lagrangian analysis of our
second-class
constrained system. The kernel of the pre-symplectic two-form,
its non-vertical part, the secondary constraint
$\Phi_{2}$, the second-order
vector field solving the field equations and tangent to the
constraint manifold are derived in detail.
Section 3 presents instead a constraint analysis within the
framework of the recently proposed generalized method.
Results and open problems are presented in section 4.
\vskip 0.3cm
\leftline {\bf 2. - Gotay-Nester Lagrangian analysis.}
\vskip 0.3cm
The Lagrangian $L$ of the model outlined in section 1
is more conveniently re-written for our purposes in the form
$$
L=16zv_{u}^{2}+2zv_{\alpha}^{2}-12zv_{u}v_{\alpha}
+2v_{u}v_{z}-v_{z}v_{\alpha}+ze^{-2\alpha}
-{z^{2}\over 4\mu}e^{-3\alpha}
\; \; \; \; ,
\eqno (2.1)
$$
where $\alpha'$, $u'$, $z'$ have been replaced by
the tangent-bundle fibre coordinates $v_{\alpha}$,
$v_{u}$, $v_{z}$ respectively.
The corresponding GN Lagrangian analysis [14] is as
follows [15]. We first
evaluate the Cartan one-form $\theta_{L} \equiv
{\partial L \over \partial v^{i}} \; dq^{i}$, and the
pre-symplectic two-form $\omega_{L} \equiv d\theta_{L}$ [16].
Given a vector field $Y$ belonging to the tangent bundle, and
setting to zero the contraction $i_{Y}\omega_{L}$, one thus
finds the kernel ${\rm ker} \; \omega_{L}$ of the
pre-symplectic two-form. Moreover, denoting by $E_{L}$ the
energy function, with corresponding one-form $dE_{L}$, the
vanishing of the contraction $i_{{\widetilde Y}_{C}}dE_{L}$ defines
the secondary constraint $\Phi_{2}$, for ${\widetilde Y}_{C}$
belonging to the non-vertical part of ${\rm ker} \; \omega_{L}$.
Denoting by $Y$ an element of ${\rm ker} \; \omega_{L}$, this
Lagrangian definition of constraints is clearly understood
acting with $i_{Y}$ on both sides of the Euler-Lagrange equations
written in the form
$$
i_{\Gamma}\omega_L+dE_{L}=0
\; \; \; \; ,
$$
and using the identity
$i_{Y}i_{\Gamma}\omega_{L}=-i_{\Gamma}i_{Y}
\omega_{L}$. This yields the condition $0=-i_{Y}dE_{L}$.
Thus, unless $i_{Y}dE_{L}$ is identically vanishing, such
calculation shows that $\Phi_{2} \equiv
i_{{\widetilde Y}_{C}}dE_{L}$ is actually the secondary
constraint of the theory (section 4).
This constraint is then preserved by requiring
that its Lie derivative along the vector field $\Gamma$ which
solves the Lagrange field equations should vanish.

Indeed, from equation (2.1) one easily finds that the Cartan
one-form and the pre-symplectic two-form are given by
$$
\theta_{L}=\Bigr(4zv_{\alpha}-12zv_{u}-v_{z}\Bigr)d\alpha
+\Bigr(32zv_{u}-12zv_{\alpha}+2v_{z}\Bigr)du+
\Bigr(2v_{u}-v_{\alpha}\Bigr)dz
\; \; \; \; ,
\eqno (2.2)
$$
$$ \eqalignno{
\omega_{L}&=\Bigr[\Bigr(4v_{\alpha}-12v_{u}\Bigr)dz+4z \; dv_{\alpha}
-12z \; dv_{u}-dv_{z}\Bigr] \wedge d\alpha \cr
&+\Bigr[\Bigr(32v_{u}-12v_{\alpha}\Bigr)dz
-12z \; dv_{\alpha}+32z \; dv_{u}+2dv_{z}
\Bigr] \wedge du \cr
&+\Bigr[2dv_{u}-dv_{\alpha}\Bigr] \wedge dz
\; \; \; \; .
&(2.3)\cr}
$$
Moreover, for a vector field $Y$ of the tangent bundle,
whose general decomposition is
$$ \eqalignno{
Y&=Y_{\alpha}(q,v){\partial \over \partial \alpha}
+Y_{u}(q,v){\partial \over \partial u}
+Y_{z}(q,v){\partial \over \partial z} \cr
&+Y_{v_{\alpha}}(q,v){\partial \over \partial v_{\alpha}}
+Y_{v_{u}}(q,v){\partial \over \partial v_{u}}
+Y_{v_{z}}(q,v){\partial \over \partial v_{z}}
\; \; \; \; ,
&(2.4)\cr}
$$
the contraction with $\omega_{L}$ is evaluated according to
the formula
$$
i_{Y}\omega_{L}={\partial^{2}L \over \partial v^{i} \partial
v^{j}}\Bigr(Y_{v}^{j} \; dq^{i}-Y_{q}^{i} \; dv^{j}\Bigr)
+{\partial^{2}L \over \partial v^{i} \partial q^{j}}
\Bigr(Y_{q}^{j} \; dq^{i}-Y_{q}^{i} \; dq^{j}\Bigr)
\; \; \; \; .
\eqno (2.5)
$$
After a lengthy calculation, this yields
$$ \eqalignno{
i_{Y}\omega_{L}&=\Bigr[4zY_{v_{\alpha}}-12zY_{v_{u}}-Y_{v_{z}}
+\Bigr(4v_{\alpha}-12v_{u}\Bigr)Y_{z}\Bigr]d\alpha \cr
&+\Bigr[-12zY_{v_{\alpha}}+32zY_{v_{u}}+2Y_{v_{z}}
+\Bigr(32v_{u}-12v_{\alpha}\Bigr)Y_{z}\Bigr]du \cr
&+\Bigr[-Y_{v_{\alpha}}+2Y_{v_{u}}+\Bigr(12v_{u}-4v_{\alpha}\Bigr)
Y_{\alpha}+\Bigr(12v_{\alpha}-32v_{u}\Bigr)Y_{u}\Bigr]dz \cr
&+\Bigr[-4zY_{\alpha}+12zY_{u}+Y_{z}\Bigr]dv_{\alpha}
+\Bigr[12zY_{\alpha}-32zY_{u}-2Y_{z}\Bigr]dv_{u} \cr
&+\Bigr(Y_{\alpha}-2Y_{u}\Bigr)dv_{z}
\; \; \; \; .
&(2.6)\cr}
$$
Thus, if we set to zero $i_{Y}\omega_{L}$, the vector
field ${\widetilde Y} \in {\rm ker} \; \omega_{L}$ is found to
take the form
$$ \eqalignno{
{\widetilde Y}&=Y_{\alpha}{\partial \over \partial \alpha}
+{1\over 2}Y_{\alpha}{\partial \over \partial u}
-2zY_{\alpha}{\partial \over \partial z}
+Y_{v_{\alpha}}{\partial \over \partial v_{\alpha}}
+\biggr[{1\over 2}Y_{v_{\alpha}}+\Bigr(2v_{u}-v_{\alpha}\Bigr)
Y_{\alpha}\biggr]{\partial \over \partial v_{u}} \cr
&+\Bigr[-2zY_{v_{\alpha}}+4zv_{\alpha}Y_{\alpha}\Bigr]
{\partial \over \partial v_{z}}
\; \; \; \; .
&(2.7)\cr}
$$
Note that, in equation (2.7), $Y_{\alpha}$ and
$Y_{v_{\alpha}}$ remain arbitrary functions of
$\alpha,u,z,v_{\alpha},v_{u}$ and $v_{z}$.
For $Y_\alpha=0$ we get the vertical kernel. For them
$dE_{L}(\widetilde Y)=0$ is identically
satisfied. By setting $Y_{v_{\alpha}}=0$
we get vector fields ${\widetilde Y}_{C}$
giving rise to constraints.
Thus, since the energy function $E_{L} \equiv
v^{i}{\partial L \over \partial v^{i}}-L$ is, in our
case, such that
$$
E_{L}=16zv_{u}^{2}+2zv_{\alpha}^{2}-12zv_{u}v_{\alpha}
+2v_{u}v_{z}-v_{z}v_{\alpha}
+{z^{2}\over 4\mu}e^{-3\alpha}-ze^{-2\alpha}
\; \; \; \; ,
\eqno (2.8)
$$
$$ \eqalignno{
dE_{L}&=\biggr(2ze^{-2\alpha}-{3z^{2}\over 4\mu}e^{-3\alpha}
\biggr)d\alpha +\biggr(2v_{\alpha}^{2}-12v_{u}v_{\alpha}
+16v_{u}^{2}-e^{-2\alpha}+{z\over 2\mu}e^{-3\alpha}\biggr)dz \cr
&+\Bigr(4zv_{\alpha}-12zv_{u}-v_{z}\Bigr)dv_{\alpha}
+\Bigr(-12zv_{\alpha}+32zv_{u}+2v_{z}\Bigr)dv_{u} \cr
&+\Bigr(2v_{u}-v_{\alpha}\Bigr)dv_{z}
\; \; \; \; ,
&(2.9)\cr}
$$
the secondary constraint $\Phi_{2}$ can be found as
$$
i_{{\widetilde Y}_{C}}dE_{L} \equiv \Phi_{2} =\Bigr(2v_{u}-
v_{\alpha}\Bigr)\Bigr(16zv_{u}-4zv_{\alpha}+2v_{z}\Bigr)
+z\biggr(4e^{-2\alpha}-{7z\over 4\mu}e^{-3\alpha}\biggr)
\; \; .
\eqno (2.10)
$$

Let us now consider a vector field $\Gamma$ which solves the
Lagrange field equations
$$
i_{\Gamma}\omega_{L}=-dE_{L}
\; \; \; \; .
\eqno (2.11)
$$
In light of equations (2.6) and (2.9), equation (2.11) yields
by comparison
$$
\Gamma_{u}=\Bigr(v_{u}-{v_{\alpha}\over 2}\Bigr)
+{1\over 2}\Gamma_{\alpha}
\; \; \; \; ,
\eqno (2.12)
$$
$$
\Gamma_{z}=2zv_{\alpha}+v_{z}-2z\Gamma_{\alpha}
\; \; \; \; ,
\eqno (2.13)
$$
$$
\Gamma_{v_{u}}={1\over 2}\Gamma_{v_{\alpha}}
+\Bigr(2v_{u}-v_{\alpha}\Bigr)\Gamma_{\alpha}
+\Bigr(2v_{\alpha}+{v_{z}\over z}\Bigr)
\Bigr({v_{\alpha}\over 2}-v_{u}\Bigr)
-{1\over 16}\biggr(8e^{-2\alpha}
-{3z\over \mu}e^{-3\alpha}\biggr)
\; \; \; \; ,
\eqno (2.14)
$$
$$
\Gamma_{v_{z}}=-2z\Gamma_{v_{\alpha}}
+4zv_{\alpha}\Gamma_{\alpha}
-2v_{\alpha}\Bigr(2zv_{\alpha}+v_{z}\Bigr)
+z\biggr(8e^{-2\alpha}-{3z\over \mu}e^{-3\alpha}\biggr)
\; \; \; \; ,
\eqno (2.15)
$$
whereas $\Gamma_{\alpha}$ and $\Gamma_{v_{\alpha}}$ remain
arbitrary functions of $\alpha$, $u$, $z$, $v_{\alpha}$,
$v_{u}$ and $v_{z}$. The dynamics is tangent to the
constraint manifold if the Lie derivative
along $\Gamma$ of the secondary constraint $\Phi_{2}$
in equation (2.10) is
vanishing, i.e. ${\cal L}_{\Gamma}\Phi_{2}=0$. This means that
the contraction $i_{\Gamma}d\Phi_{2}$ should vanish, and
leads to a restriction on the coefficient $\Gamma_{\alpha}$,
which is found to take the value ${\widetilde \Gamma}_{\alpha}$
such that
$$
{\widetilde \Gamma}_{\alpha}={\Bigr(2zv_{\alpha}+v_{z}\Bigr)
\biggr[{11z\over 4\mu}e^{-3\alpha}-2e^{-2\alpha}+
2\Bigr(2v_{u}-v_{\alpha}\Bigr)
\Bigr(8v_{u}-2v_{\alpha}+{v_{z}\over z}\Bigr)\biggr]
\over
\biggr[{49z^{2}\over 4\mu}e^{-3\alpha}-16ze^{-2\alpha}
+4z\Bigr(2v_{u}-v_{\alpha}\Bigr)
\Bigr(8v_{u}-2v_{\alpha}+{v_{z}\over z}\Bigr)\biggr]}
\; \; \; \; ,
\eqno (2.16)
$$
since the one-form $d\Phi_{2}$ is
$$ \eqalignno{
d\Phi_{2}&=\biggr({21z^{2}\over 4\mu}e^{-3\alpha}-8ze^{-2\alpha}
\biggr)d\alpha
+\biggr(4v_{\alpha}^{2}-24v_{u}v_{\alpha}+32v_{u}^{2}
+4e^{-2\alpha}-{7z\over 2\mu}e^{-3\alpha}\biggr)dz\cr
&+\Bigr(8zv_{\alpha}-24zv_{u}-2v_{z}\Bigr)dv_{\alpha}
+\Bigr(-24zv_{\alpha}+64zv_{u}+4v_{z}\Bigr)dv_{u}\cr
&+\Bigr(4v_{u}-2v_{\alpha}\Bigr)dv_{z}
\; \; \; \; ,
&(2.17)\cr}
$$
and the various coefficients of $\Gamma_{v_{\alpha}}$
appearing in $i_{\Gamma}d\Phi_{2}$ add up to zero.
The geometrical meaning of our calculation is as follows.
If $\Gamma$ solves the Lagrange equations (2.11), for any
vector field ${\widetilde Y} \in {\rm ker} \; \omega_{L}$ one
finds
$$
i_{\Gamma + {\widetilde Y}} \; \omega_{L}
=\Bigr(i_{\Gamma} \; \omega_{L}\Bigr)
+\Bigr(i_{\widetilde Y} \; \omega_{L}\Bigr)
=-dE_{L}
\; \; \; \; .
\eqno (2.18)
$$
In other words, the arbitrariness of $\Gamma_{\alpha}$ and
$\Gamma_{v_{\alpha}}$ reflects the possibility of adding to
any solution of equation (2.11) an arbitrary vector field
${\widetilde Y} \in {\rm ker} \; \omega_{L}$. However, if
the vector field $\Gamma$ is also tangent to the constraint
manifold, only $\Gamma_{v_{\alpha}}$ remains arbitrary.
The vector field $\Gamma_{T}$ that solves equation (2.11)
and is tangent to the constraint manifold is thus found
to be
$$
\Gamma_{T}={\widetilde \Gamma}_{\alpha}{\partial \over
\partial \alpha}+{\widetilde \Gamma}_{u}
{\partial \over \partial u}
+{\widetilde \Gamma}_{z}{\partial \over \partial z}
+\Gamma_{v_{\alpha}}{\partial \over \partial v_{\alpha}}
+{\widetilde \Gamma}_{v_{u}}{\partial \over \partial v_{u}}
+{\widetilde \Gamma}_{v_{z}}{\partial \over \partial v_{z}}
\; \; \; \; ,
\eqno (2.19)
$$
where ${\widetilde \Gamma}_{\alpha}$ has been evaluated as in
equation (2.16), and ${\widetilde \Gamma}_{u}$,
${\widetilde \Gamma}_{z}$, ${\widetilde \Gamma}_{v_{u}}$,
${\widetilde \Gamma}_{v_{z}}$ are values taken by the
right-hand sides of equations (2.12)-(2.15) at
$\Gamma_{\alpha}={\widetilde \Gamma}_{\alpha}$. The
arbitrariness of $\Gamma_{v_{\alpha}}$ is due to the existence
of a vertical kernel.
\vskip 0.3cm
\leftline {\bf 3. - Generalized method.}
\vskip 0.3cm
We here study a different and more recent
method for the constrained analysis of a
dynamical system. Interestingly, it does not rely on a Cartan
one-form or a pre-symplectic two-form as the method studied
in section 2, but enables one to derive constraints by
looking directly at the equations of motion, and can be applied
to any system of differential equations in implicit form.

To describe the method, let us assume that a system of
implicit dynamical equations is given in the particular form
$$
a_{ij}(q,v){\dot v}^{i}-f_{j}(q,v)=0
\; \; \; \; ,
\eqno (3.1)
$$
$$
b_{ij}(q,v)\Bigr({\dot q}^{i}-v^{i}\Bigr)=0
\; \; \; \; .
\eqno (3.2)
$$
One now has a choice of first- or second-order formalism.
If first-order theory is used, the time-derivatives of
the positions $q^{i}$ are not identified with the
velocities $v^{i}$. Denoting by $\psi^{j}$ a vector such
that $\psi^{j}b_{ij}=0$, the most general form of Eq. (3.2)
is
$$
b_{ij}(q,v)\Bigr({\dot q}^{i}-v^{i}+\psi^{i}\Bigr)=0
\; \; \; \; .
\eqno (3.3)
$$
Thus, in first-order theory, one finds
$$
{\dot q}^{i}=v^{i}-\psi^{i}
\; \; \; \; .
\eqno (3.4)
$$
The constraints are found by solving
for $\phi^{j}$ the equation
$\phi^{j}a_{ij}=0$. This leads to
a compatibility condition of the kind
$\phi^{j}f_{j}=0$. If this equation is not identically
satisfied then $\phi^{j}f_{j}$ is the first-generation
constraint of the theory.
The constraints are preserved by requiring that
$$
{d\over dt}\Bigr(\phi^{j}f_{j}\Bigr)
={\partial \Bigr(\phi^{j}f_{j}\Bigr) \over \partial
q^{k}}{\dot q}^{k}+{\partial \Bigr(\phi^{j}f_{j}\Bigr)
\over \partial v^{k}}{\dot v}^{k}=0
\; \; \; \; .
\eqno (3.5)
$$
In second-order theory one has instead
${\dot q}^{i}=v^{i}$, hence $\psi^{i}=0$ in (3.4).
Note that the arguments
developed so far do not rely on the existence of a Lagrangian,
in agreement with what we said.

However, if a Lagrangian
is known for the model under consideration, $a_{ij}$ is
the Hessian matrix $H_{ij}$, whereas
$
b_{ij}={\partial^{2}L \over \partial q^{i} \partial
{\dot q}^{j}}-{\partial^{2}L \over \partial q^{j}
\partial {\dot q}^{i}}
$.
Multiplying by a vector $A^{i}$ the second-order Lagrange equations
$$
{d\over dt}{\partial L \over \partial v^{i}}
-{\partial L \over \partial q^{i}}=0
\; \; \; \; ,
\eqno (3.6)
$$
where
$$
v^{i}(t)={d\over dt}q^{i}(t)
\; \; \; \; ,
\eqno (3.7)
$$
one finds
$$
{\partial^{2}L \over \partial v^{i} \partial q^{j}}
v^{j}A^{i}-{\partial L \over \partial q^{i}}A^{i}=0
\; \; \; \; ,
\eqno (3.8)
$$
which is a restriction on Cauchy data, i.e. a constraint.
Following [17-18], such constraints are called
{\it first-generation} constraints. Clearly, the number of
independent eigenvectors corresponding to zero eigenvalues
of $H_{ij}$ is $N-K$, where $K$ is the maximal rank of the
Hessian matrix. Hence first-generation constraints can be
denoted by $\psi_{m}^{{\widehat {\rm I}}}$,
$m=1,2,...,M \leq N-K$. Second-generation constraints
(if any) are then obtained by requiring that the evolution
of the system should be tangent to the first-generation
constraint manifold, i.e.
$$
{d\over dt} \psi^{{\widehat {\rm I}}}
\equiv \psi^{{\widehat {\rm
II}}}=0
\; \; \; \; .
\eqno (3.9)
$$
Analogously, third-generation constraints are determined by
imposing that the evolution should be tangent to the
manifold defined by first- and second-generation
constraints, and so on.

A similar analysis can be repeated for first-order
Lagrange equations. This leads to
$$
{\partial^{2}L\over\partial v^{i}\partial
v^{j}}v^{j}A_{q}^{i}+\left({\partial L\over\partial v^{i}}-
{\partial^{2}L\over\partial q^{i}\partial v^{j}}v^{j}
\right)A_{v}^{i}=0
\; \; \; \; .
\eqno (3.10)
$$
A first-generation constraint has been thus obtained within
the first-order formalism. All further constraints
(i.e. second-generation, third-generation and so on) are
then found by imposing that the evolution of the system
should be tangent to the constraint manifold.

We now apply this new method to our cosmological model.
The first-order equations of motion are
$$
2\Bigr({\dot u}-v_{u}\Bigr)-
\Bigr({\dot \alpha}-v_{\alpha}\Bigr)=0
\; \; \; \; ,
\eqno (3.11)
$$
$$
2z\Bigr({\dot z}-v_{z}\Bigr)-
12z\Bigr({\dot \alpha}-v_{\alpha}\Bigr)+
32z\Bigr({\dot u}-v_{u}\Bigr)=0
\; \; \; \; ,
\eqno (3.12)
$$
$$
4z\Bigr({\dot\alpha}-v_{\alpha}\Bigr)-
12z\Bigr({\dot u}-v_{u}\Bigr)-
\Bigr({\dot z}-v_{z}\Bigr)=0
\; \; \; \; ,
\eqno (3.13)
$$
$$
{\dot v_{z}}-6{\dot z}v_{\alpha}-6z{\dot v}_{\alpha}+
16{\dot z}v_{u}+16{\dot v}_{u}z=0
\; \; \; \; ,
\eqno (3.14)
$$
$$
-{\dot v}_{z}+4{\dot z}v_{\alpha}+4z{\dot v}_{\alpha}-
12z{\dot v}_{u}-12{\dot z}v_{u}+2ze^{-2\alpha}-{3\over4}
{z^{2}\over\mu}e^{-3\alpha}=0
\; \; \; \; ,
\eqno (3.15)
$$
$$ \eqalignno{
-{\dot v}_{\alpha}+2{\dot v}_{u}-e^{-2\alpha}
-2v^{2}_{\alpha}&
+12v_{u}v_{\alpha}-16v^{2}_{u}+
{z\over 2\mu}e^{-3\alpha}\cr
&=\Bigr(4v_{\alpha}-
12v_{u}\Bigr)\Bigr({\dot \alpha}-v_{\alpha}\Bigr)\cr
&+\Bigr(32v_{u}-12v_{\alpha}\Bigr)\Bigr({\dot u}-
v_{u}\Bigr)
\; \; \; \; .
&(3.16)\cr}
$$
Since the Hessian matrix $H_{ij}$ is of rank 2 in our case,
its kernel is one-dimensional. The vector of the kernel
of $H_{ij}$ is found to be
$$
A^{i}=\pmatrix {2\cr 1\cr -4z\cr}
\; \; \; \; .
\eqno (3.17)
$$
The equations of motion (3.14)-(3.16) can be seen as a row
vector which, multiplied by $A^{i}$, yields the
first-generation constraint. Such a constraint is thus
defined by
$$
\psi^{\widehat {\rm I}}\equiv
A^{i}{\partial^{2}L\over\partial v^{i}
\partial q^{j}}{\dot q}^{j}-
A^{i}{\partial L\over \partial q^{i}}-
A^{i}{\partial^{2}L\over \partial v^{j}
\partial q^{i}}\Bigr({\dot q}^{j}-
v^{j}\Bigr)=0
\; \; \; \; .
\eqno (3.18)
$$
Bearing in mind equations (3.11)-(3.13), one finds
$$
\psi^{\widehat {\rm I}}=\Bigr(2v_{u}-
v_{\alpha}\Bigr)\Bigr(8zv_{u}-2zv_{\alpha}+v_{z}\Bigr)
+z\biggr(2e^{-2\alpha}-{7z\over 8\mu}e^{-3\alpha}\biggr)
\; \; \; \; .
\eqno (3.19)
$$
The constraint
$\psi^{\widehat {\rm I}}$ is preserved if
${d\over dt}\psi^{\widehat {\rm I}}=0$, which implies
$$
{\dot \alpha}={\Bigr(2zv_{\alpha}+v_{z}\Bigr)
\biggr[{11z\over 4\mu}e^{-3\alpha}-2e^{-2\alpha}+
2\Bigr(2v_{u}-v_{\alpha}\Bigr)
\Bigr(8v_{u}-2v_{\alpha}+{v_{z}\over z}\Bigr)\biggr]
\over
\biggr[{49z^{2}\over 4\mu}e^{-3\alpha}-16ze^{-2\alpha}
+4z\Bigr(2v_{u}-v_{\alpha}\Bigr)
\Bigr(8v_{u}-2v_{\alpha}+{v_{z}\over z}\Bigr)\biggr]}
\; \; \; \; .
\eqno (3.20)
$$
By imposing that the evolution of the system should be
tangent to the constraint manifold we have derived a
formula for $\dot \alpha$, which is the analogue of
${\widetilde \Gamma}_{\alpha}$ in the Gotay-Nester analysis of
section 2 (Eq. (2.16)). Thus,
by using the generalized method, the
constraint analysis is completed once we determine
$\dot \alpha$, which remains arbitrary if one looks simply
at the equations of motion which govern the dynamics.

It is now very instructive to perform the constraint
analysis within the second-order formalism. The second-order
Lagrange equations (3.6)-(3.7) read
$$
v_{\alpha}={\dot \alpha}
\; \; \; \; ,
\eqno (3.21)
$$
$$
v_{u}={\dot u}
\; \; \; \; ,
\eqno (3.22)
$$
$$
v_{z}={\dot z}
\; \; \; \; ,
\eqno (3.23)
$$
$$
{\dot v_{z}}-6v_{z}v_{\alpha}-6z{\dot v}_{\alpha}+
16v_{z}v_{u}+16{\dot v}_{u}z=0
\; \; \; \; ,
\eqno (3.24)
$$
$$
-{\dot v}_{z}+4v_{z}v_{\alpha}+4z{\dot v}_{\alpha}-
12z{\dot v}_{u}-12v_{z}v_{u}+2ze^{-2\alpha}-{3\over4}
{z^{2}\over\mu}e^{-3\alpha}=0
\; \; \; \; ,
\eqno (3.25)
$$
$$
-{\dot v}_{\alpha}+2{\dot v}_{u}-e^{-2\alpha}
-2v^{2}_{\alpha}
+12v_{u}v_{\alpha}-16v^{2}_{u}+
{z\over 2\mu}e^{-3\alpha}=0
\; \; \; \; .
\eqno (3.26)
$$
The kernel of the Hessian matrix is again given by (3.17),
and the first-generation constraint (3.8) turns out to
coincide with the first-generation constraint (3.19). The
search for second-generation constraints leads to
$$
\eqalignno {
\psi^{\widehat {\rm II}}&
\equiv {d\over dt}\psi^{\widehat {\rm I}}\cr
&={\dot \alpha}\biggr[{49z^{2}\over 4\mu}e^{-3\alpha}
-16ze^{-2\alpha}
+4z\Bigr(2{\dot u}-{\dot \alpha}\Bigr)
\Bigr(8{\dot u}-2{\dot \alpha}
+{{\dot z}\over z}\Bigr)\biggr] \cr
&-\Bigr(2z{\dot \alpha}+{\dot z}\Bigr)
\biggr[{11z\over 4\mu}e^{-3\alpha}
-2e^{-2\alpha}+
2\Bigr(2{\dot u}-{\dot \alpha}\Bigr)
\Bigr(8{\dot u}-2{\dot \alpha}
+{{\dot z}\over z}\Bigr)\biggr]
\; \; \; \; .
&(3.27)\cr}
$$
Requiring that the evolution of the system should be
tangent to the constraint manifold, and defining
$$
{\widetilde D}\equiv 16ze^{-2\alpha}-{49\over 4\mu}z^{2}e^{-3\alpha}
+4z\Bigr({\dot \alpha}-2{\dot u}\Bigr)
\Bigr(8{\dot u}-2{\dot \alpha}+
{{\dot z}\over z}\Bigr)
\; \; \; \; ,
\eqno (3.28)
$$
one finds
$$
\eqalignno {
{\widetilde D}{\dot \alpha}&=
{\dot \alpha}\biggr[{e^{-3\alpha}\over 4\mu}
\Bigr(-123z^{2}{\dot \alpha}+98z{\dot z}
-48z^{2}{\dot u}\Bigr)\cr
&+e^{-2\alpha}\Bigr(16z{\dot \alpha}
-16{\dot z}+32z{\dot u}\Bigr)\cr
&-{z\over 2}\Bigr(8{\dot u}-2{\dot \alpha}
+{{\dot z}\over z}\Bigr)
\biggr(8e^{-2\alpha}-{3z\over \mu}e^{-3\alpha}\biggr)\cr
&+4\Bigr(2{\dot u}-{\dot \alpha}\Bigr)
\biggr(\Bigr(2{\dot \alpha}-8{\dot u}\Bigr){\dot z}
-{{\dot z}^{2}\over z}
\biggr)\biggr]\cr
&-z\biggr(8e^{-2\alpha}-{3z\over \mu}e^{-3\alpha}\biggr)
\biggr[{11z\over 4\mu}e^{-3\alpha}-2e^{-2\alpha}
+2\Bigr(2{\dot u}-{\dot \alpha}\Bigr)
\Bigr(8{\dot u}-2{\dot \alpha}+{{\dot z}\over z}\Bigr)
\biggr]\cr
&-4\Bigr(2z{\dot \alpha}+{\dot z}\Bigr)
\Bigr({\dot \alpha}-2{\dot u}\Bigr)
\Bigr(8{\dot u}-2{\dot \alpha}+{{\dot z}\over z}\Bigr)\cr
&+{e^{-3\alpha}\over 4\mu}\Bigr(2z{\dot \alpha}+{\dot z}\Bigr)
\Bigr(27z{\dot \alpha}-14{\dot z}\Bigr)\cr
&+2e^{-2\alpha}\Bigr(2z{\dot \alpha}
+{\dot z}\Bigr){{\dot z}\over z}
\; \; \; \; .
&(3.29)\cr}
$$
Note that a substantial difference occurs with respect to a
first-order formalism, since we find a second-generation
constraint which is absent in the first-order case. Hence the
variable $v_{\alpha}$ is determined, which remains instead
arbitrary in the first-order analysis.

The occurrence of the second-generation constraint, which has
no equivalent in Dirac's Hamiltonian treatment [2] of section
1, is not a peculiar property  of the {\it generalized method}
of this section, but rather of the second-order formalism. A
second-order Gotay-Nester analysis yields the same result.
Indeed, for any point Lagrangian in a second-order theory
with constraints, the contraction of the energy one-form with
a vector field $A$ in the kernel of the pre-symplectic two-form
yields the constraint
$$
\phi(q,{\dot q}) \equiv A^{s} \left[{\dot q}^{h}
{\partial^{2}L\over \partial {\dot q}^{s}\partial q^{h}}
-{\partial L \over \partial q^{s}}\right]
\; \; \; \; ,
\eqno (3.30)
$$
which coincides with the second-order
first-generation constraint (3.7)-(3.8). The Lie derivative of
$\phi(q,{\dot q})$ along the second-order vector field
$$
\Gamma={\dot q}^{i}{\partial \over \partial q^{i}}
+\Gamma_{{\dot q}}^{i}{\partial \over \partial
{\dot q}^{i}}
\; \; \; \; ,
\eqno (3.31)
$$
leads to a further constraint which, in our paper, coincides
(by construction) with the second-generation constraint (3.27).

Indeed, our model is not the only case of point Lagrangian
where second-order formalism leads to further constraints
with respect to first-order formalism. For example, if one
studies the point Lagrangian (cf. [19])
$$
L \equiv {1\over 2}v_{1}^{2}+v_{1}q_{2}
+\Bigr(1-\alpha\Bigr)v_{2}q_{1}
+{\beta \over 2}{\Bigr(q_{1}-q_{2}\Bigr)}^{2}
\; \; \; \; ,
\eqno (3.32)
$$
with $\alpha^{2}-\beta \not =0$, first-order formalism
only leads to the constraint
$$
\Phi_{2} \equiv -\alpha v_{1}+\beta \Bigr(q_{1}-q_{2}\Bigr)
\; \; \; \; ,
\eqno (3.33)
$$
whereas second-order formalism also leads to the further
constraint
$$
\Phi_{3} \equiv \Bigr(\alpha^{2}-\beta \Bigr){\dot q}_{2}
+\beta {\dot q}_{1}-\alpha \beta \Bigr(q_{1}-q_{2}\Bigr)
\; \; \; \; .
\eqno (3.34)
$$
In the latter case, the vector field solving the Lagrange
field equations (2.11) takes the form
$$
\Gamma = {\dot q}_{1} {\partial \over \partial q_{1}}
+{\dot q}_{2}{\partial \over \partial q_{2}}
+\biggr(-\alpha {\dot q}_{2}+\beta \Bigr(q_{1}-q_{2}\Bigr)\biggr)
{\partial \over \partial {\dot q}_{1}}
\; \; \; \; .
\eqno (3.35)
$$
The additional constraint (3.34) is obtained since velocities
have been taken to be time-derivatives of positions.

A further example of the inequivalence of first- and
second-order theory is given by the following Lagrangian
on the tangent bundle of $SU(2)$:
$$
L \equiv {\rm Tr}\Bigr(\sigma_{3}S^{-1}{\dot S}\Bigr)
\; \; \; \; ,
\eqno (3.36)
$$
where $S \in SU(2)$.
The corresponding pre-symplectic 2-form is found to be
$$
\omega_{L}=-{\rm Tr} \; \sigma_{3}\Bigr(S^{-1}dS
\wedge S^{-1}dS \Bigr)
\; \; \; \; ,
\eqno (3.37)
$$
and the first-order dynamics is given by the vector field
$X_{3}$ associated with the one-parameter subgroup
$e^{{it \over 2}\sigma_{3}}$. In second-order
formalism, however, there is {\it no dynamics} compatible
with the Lagrangian (3.36). This Lagrangian is a Chern-Simons
Term, and hence is relevant for modern field theory.

Interestingly, we may use the constraints (3.19) and (3.27)
to derive physical predictions. In other words, by imposing
the first-generation constraint $\psi^{\widehat {\rm I}}=0$
we may cast the second-generation constraint in the form
$$
\psi^{\widehat {\rm II}}
={\dot \alpha}\biggr({27 z^{2}\over 4\mu}e^{-3\alpha}
-12ze^{-2\alpha}\biggr)
+{\dot z}\biggr(6e^{-2\alpha}-{9z\over 2\mu}e^{-3\alpha}
\biggr)=0
\; \; \; \; .
\eqno (3.38)
$$
Equation (3.38) leads to
$$
{dz \over z}=
{{\biggr(4e^{\alpha}-{9z\over 4\mu}\biggr)} \over
{\biggr(2e^{\alpha}-{3z\over 2\mu}\biggr)}}
d\alpha
\; \; \; \; .
\eqno (3.39)
$$
Thus, defining
$$
x \equiv e^{-\alpha}
\; \; \; \; ,
\eqno (3.40)
$$
$$
xz \equiv \eta(\xi(x))
\; \; \; \; ,
\eqno (3.41)
$$
$$
\xi(x) \equiv \log(x)
\; \; \; \; ,
\eqno (3.42)
$$
$$
f(\eta) \equiv
-{{\biggr(4-{9\eta\over 4\mu}\biggr)} \over
{\biggr(2-{3\eta\over 2\mu}\biggr)}}
\; \; \; \; ,
\eqno (3.43)
$$
differentiation of (3.41) with respect to $x$
and comparison with (3.39) leads to the differential
equation
$$
{d\eta \over d\xi}=\eta \Bigr(1+f(\eta)\Bigr)
\; \; \; \; .
\eqno (3.44)
$$
The corresponding solution for $x$ may be expressed in the
form (see Eq. I.I26 on page 316 of [20])
$$
\log(x)=\int_{C}^{xz}{d\eta \over \eta(1+f(\eta))}
\; \; \; \; .
\eqno (3.45)
$$
By virtue of (3.43), Eq. (3.45) implies, after performing
some standard integrals, that
$$
x^{2}z \biggr({3\over 4\mu}xz-2 \biggr)={\widetilde A}
\; \; \; \; ,
\eqno (3.46)
$$
where ${\widetilde A}$
is an integration constant. Thus, by using the
definition (3.40) and the relation between $z$ and the
scalar curvature: $z={\mu \over 3}e^{3\alpha}R$, (3.46)
leads to the second-order algebraic equation for the scalar
curvature
$$
a^{3}R^{2}-8aR-4{\widetilde A}=0
\; \; \; \; .
\eqno (3.47)
$$
Although it seems impossible
to solve by analytic methods the field equations in Hamiltonian
(cf. [2]) or Lagrangian form, we have been able to re-express
first- and second-generation constraints in terms of physically
relevant quantities, i.e. the cosmic scale factor and the
scalar curvature. Interestingly, the roots of (3.47) are
given by
$$
R={4\over a^{2}} \pm {2\over a^{2}}
\sqrt{4+{\widetilde A}a}
\; \; \; \; .
\eqno (3.48)
$$
Thus, if second-order formalism is used, the scalar curvature
does not vanish in vacuum in the presence of torsion, unless the
constant ${\widetilde A}$
is set to zero and the negative sign is chosen
in front of the square root. Moreover,
if the constant ${\widetilde A}$
is negative, the cosmic scale factor is
bound to remain less than or equal to
${4\over {\mid {\widetilde A} \mid}}$.

By contrast, if torsion vanishes, our model corresponds to
a Hamiltonian system with first-class constraints (cf. [7]).
At a Lagrangian level, the first-generation constraint is
given by ${E_{L}/N}$, where $E_{L}$ is the energy function and
$N$ is the lapse function. Thus, such a constraint vanishes if
and only if the energy function vanishes, and is automatically
preserved, since $E_{L}$ is constant along solutions of the field
equations (hence its Lie derivative along the vector field solving
the Lagrange equations vanishes).
\vskip 0.3cm
\leftline {\bf 4. - Concluding remarks.}
\vskip 0.3cm
The contribution of this paper is a detailed
application of the generalized method of section 3.
Since it relies on an approach [17-18] whose
range of applicability is wider than
any previous (Lagrangian) method, we found it appropriate
to focus on this technique in our paper.
Remarkably, in our specific model, if Lagrange equations
are studied in second-order formalism, a second-generation
constraint is found which is absent in first-order formalism.
This constraint has been expressed in terms of the physical
quantities of the problem, as shown in Eq. (3.38).
One thus finds the equation (3.47) for the scalar curvature,
whose roots are given by (3.48). It turns out that the scalar
curvature may not vanish in vacuum.
Interestingly, an upper limit for
the cosmic scale factor exists if the
constant ${\widetilde A}$ in (3.48)
is negative. Hence second-order theory is found to yield
relevant information about the early universe in our toy model.

Note also that the secondary constraint of equation
(2.10) is obtained multiplying
by $-{1\over 2}$ the secondary constraint of equation (1.7), as one may
check by using (2.1) and the map
$$
FL(q^i,v^i)\equiv\left(q^i,p_i\equiv{\partial
L\over\partial v^i}\right)
\; \; \; \; ,
\eqno (4.1)
$$
which expresses the map in local
coordinates from the tangent bundle $TQ$ to the cotangent bundle $T^*Q$.
This implies that, up to an unessential
proportionality constant, the Hamiltonian and the Lagrangian
method lead to the same secondary constraint. However, the
careful reader may have noticed that, using tangent-bundle
formalism, no primary constraint occurs.
Such property is not surprising for the following reasons.
If ${\rm det} \mid {\partial^{2}L \over \partial
v^{i} \partial v^{j}} \mid=0$, the map $FL$ defined in
(4.1) maps the tangent bundle $TQ$ to the
primary-constraint submanifold ${\widehat \Sigma}$ of the
phase space $T^{*}Q$. Thus, $(FL)^{*}$ maps (by definition
of pull-back) forms on ${\widehat \Sigma}$ to forms on
$TQ$, and the forms $\omega_{L}$, $dE_{L}$ appearing in the
Lagrange equation (2.11) correspond to geometrical objects
{\it already} living on ${\widehat \Sigma} \subset
T^{*}Q$. Interestingly, the first set of constraints one
actually evaluates in the Lagrangian formalism are the
equivalent of secondary constraints of Dirac's theory.
This property appears very important for first-class
theories such as Maxwell's electrodynamics and GR: the
Lagrangian method is a powerful way to evaluate directly
the constraints which govern the theory, i.e. first-class
secondary constraints. Although the question remains open
[11-13,21],
Dirac's consideration of first-class
secondary constraints appears more natural within the
Lagrangian approach.

A naturally occurring question is whether a Lagrangian multisymplectic
analysis [22] of first-class theories such as Einstein's GR
offers any advantage with respect to the
ADM Hamiltonian method [13,23-24] which relies instead on the
3+1 split of the space-time geometry,
especially in light of the longstanding
unsolved problems of the quantization program [13].
Work is now in progress by the authors on some of these issues,
and we hope that the geometrical reformulation of certain
properties of classical field theory presented in this paper will serve
as a first step towards this more ambitious goal.
\vskip 0.1cm
\centerline {* * *}
\vskip 0.1cm
\noindent
G. Esposito is grateful to Professor Abdus Salam, the
International Atomic Energy Agency and UNESCO for hospitality
at the International Centre for Theoretical Physics, and to
Professor Dennis Sciama for hospitality at the SISSA of
Trieste, during the early stages of this work.
G. Marmo is indebted to Professor Ibort for hospitality at
the Departamento de Fisica Teorica of the Universidad
Complutense of Madrid.
Conversations with Giovanna Mendella and correspondence
with Claes Uggla are also gratefully acknowledged.
\vskip 0.1cm
\leftline {REFERENCES}
\vskip 0.1cm

[1] G. MORANDI, C. FERRARIO, G. LO VECCHIO, G. MARMO and
C. RUBANO: {\it Phys. Reports}, {\bf 188}, 147 (1990).

[2] G. ESPOSITO: {\it Nuovo Cimento B}, {\bf 104}, 199 (1989).

[3] E. SEZGIN and P. van NIEUWENHUIZEN: {\it Phys. Rev. D}, {\bf 21},
3269 (1980).

[4] I. A. NIKOLIC: {\it Phys. Rev. D}, {\bf 30}, 2508 (1984).

[5] V. SZCZYRBA: {\it Phys. Rev. D}, {\bf 36}, 351 (1987).

[6] V. SZCZYRBA: {\it J. Math. Phys.} (N.Y.), {\bf 28},
146 (1987).

[7] G. T. HOROWITZ: {\it Phys. Rev. D}, {\bf 31}, 1169 (1985).

[8] G. ESPOSITO: {\it Nuovo Cimento B}, {\bf 106}, 1315 (1991).

[9] M. A. H. MACCALLUM: {\it Anisotropic and inhomogeneous
relativistic cosmologies}, in {\it General Relativity, an
Einstein Centenary Survey}, edited by S. W. Hawking and
W. Israel (Cambridge University Press, Cambridge, 1979).

[10] D. BOULWARE: {\it Quantization of higher-derivative theories
of gravity}, in {\it Quantum Theory of Gravity}, edited by S. M.
Christensen (Adam Hilger, Bristol, 1984).

[11] P. A. M. DIRAC: {\it Lectures on Quantum Mechanics},
Belfer Graduate School of Science (Yeshiva University,
New York, 1964).

[12] G. MARMO, N. MUKUNDA and J. SAMUEL: {\it Riv. Nuovo Cimento},
{\bf 6}, 1 (1983).

[13] G. ESPOSITO: {\it Quantum Gravity, Quantum Cosmology and
Lorentzian Geometries}, second corrected and enlarged edition,
Lecture Notes in Physics, New Series
m: Monographs, Volume m12 (Springer-Verlag, Berlin, 1994).

[14] M. J. GOTAY and J. M. NESTER: {\it Ann. Inst. H. Poincar\'e A},
{\bf 30}, 129 (1979).

[15] J. F. CARINENA, C. LOPEZ and N. ROMAN-ROY:
{\it J. Geom. Phys.}, {\bf 4}, 315 (1987).

[16] G. MARMO, E. J. SALETAN, A. SIMONI and B. VITALE:
{\it Dynamical Systems, a Differential Geometric Approach to
Symmetry and Reduction} (John Wiley, New York, 1985).

[17] G. MENDELLA: {\it Geometrical Aspects of Lagrangian
Dynamical Systems with Dirac Constraints},
Physics-Degree Thesis (University of Napoli, 1987);
{\it A Geometrical Formalism for Constraints'
Theory}, Ph.D. Thesis (University of Napoli, 1992).

[18] G. MARMO, G. MENDELLA and W. M. TULCZYJEW:
{\it Integrability of Implicit Differential Equations},
paper in preparation (DSF preprint 93/51).

[19] K. SUNDERMEYER: {\it Constrained Dynamics}
(Springer-Verlag, Berlin, 1982).

[20] E. KAMKE: {\it Differentialgleichungen
L\"{o}sungsmethoden und
L\"{o}sungen} (Chelsea Publishing Company, New York, 1959).

[21] M. E. V. COSTA, H. D. GIROTTI and T. J. M. SIMOES:
{\it Phys. Rev. D}, {\bf 32}, 405 (1985).

[22] M. GOTAY, J. ISENBERG, J. MARSDEN, R. MONTGOMERY,
J. SNIATYCKI and P. B. YASSKIN: {\it Momentum Maps and the
Hamiltonian Treatment of Classical Field Theories with
Constraints}, to appear in Mathematical Sciences Research
Institute Publications (Springer-Verlag, Berlin).

[23] A. ASHTEKAR: {\it Lectures on Non-Perturbative Canonical
Gravity} (World Scientific, Singapore, 1991).

[24] M. FERRARIS, M. FRANCAVIGLIA and I. SINICCO: {\it Nuovo Cimento B},
{\bf 107}, 1303 (1992).

\bye